\documentclass[12pt]{article}
\usepackage{xcolor, soul}
\sethlcolor{yellow}
\usepackage{amsmath}
\usepackage{amssymb}
\begin{document}
\vskip 2cm
\begin{center}
{\bf {\Large  3D Jackiw-Pi  Model: (Anti-)Chiral Superfield Approach to BRST Formalism}}\\

\vskip 3.2cm

{\sf B. Chauhan$^{(a)}$}\\ \vskip 0.4 cm
$^{(a)}$ {\it Physics Department,  Centre of Advance Studies, Institute of Science,}\\
{\it Banaras Hindu University, Varanasi - 221 005, (U.P.), India}\\\vskip 0.4cm
{\small {\sf {E-mails: bchauhan501@gmail.com}}}
\end{center}

\vskip 3cm

\noindent
{\bf Abstract:}
 We discuss and derive  the  continuous  Becchi-Rouet-Stora-Tyutin (BRST)and  
 anti-BRST symmetry transformations for the Jackiw-Pi (JP) model of  three (2 + 1)-dimensional  (3D)
massive non-Abelian 1-form gauge theory by exploiting the standard technique of (anti-)chiral 
superfield approach (ACSA) to  BRST formalism where a few appropriate and specific sets of 
(anti-)BRST invariant quantities (i. e. physical quantities at quantum level) play a very important role. 
We provide the explicit derivation of  the nilpotency and absolute  anticommutativity properties of (anti-)BRST conserved charges and 
existence of Curci-Ferrari  (CF)-condition within the realm  of ACSA to BRST formalism where we take only a single Grassmannian
variable into account. We also provide the clear proof of (anti-) BRST invariances of the coupled (but equivalent) Lagrangian densities
within the framework of ACSA to BRST approach  where the emergence  of the CF-condition is observed. \\

\vskip 1.2cm
\noindent
PACS numbers: 03.70.+k, 11.15.-q,  11.10.Kk, 12.90.+b \\

\vskip 0.5 cm 
\noindent
{\it {Keywords}}: 3D Jackiw-Pi model; massive non-Abelian 1-form  gauge theory;
(anti-)BRST symmetries; Curci-Ferrari condition; (anti-)chiral superfield approach;
 nilpotency and absolute anticommutativity; (anti-)BRST conserved charges

\newpage
\noindent
\section{Introduction}

Gauge theories describe the dynamics of the elementary particles under local
transformations according to certain operations which   is  important theory  because many
 successful field theories explaining   three (i.e. electromagnetic, strong, weak)
out of four fundamental interactions of nature  can be theoretically described. 
A gauge theory is endowed with the  first-class constraints in the terminology  
of Dirac's prescription for the classification scheme of constraints [1, 2].
The key signatures of a gauge invariant theory are the 
invariance and singularity of the Lagrangian density  for a given theory. 
Quantum electrodynamics is an example of  Abelian gauge theory with the symmetry group $U(1)$ which  has one gauge field
(i.e.  the electromagnetic four-potential) with the photon being the gauge boson. 
The standard model of particle physics  is based on the  non-Abelian
gauge theory with the symmetry group $U(1) \times SU(2) \times SU(3)$ which  has twelve gauge bosons (i.e. a photon,
three weak bosons and eight gluons). The non-Abelian gauge theory is very important in the different prospectives of field theories as
this theory is the foundation of electroweak and strong interactions. Our present investigation
 is based on the  massive non-Abelian 1-form gauge theory.

The covariant quantization of a gauge-field system has a long-time history 
started from the famous works of Feynman [3], Faddeev  and Popov [4] and DeWitt [5].
For the covariant canonical quantization of a given gauge invariant theory, the Becchi-Rouet-Stora-Tyutin (BRST) 
formalism  plays an important role where infinitesimal local gauge parameter is replaced by the ghost and anti-ghost fields [6-9]
to respect the unitarity of the given theory.
Thus, we have two supersymmetric-type  global BRST  $(s_b)$ and anti-BRST $(s_{ab})$  symmetry  transformations
at the quantum level. These symmetry transformations have two important and key features.
 Mathematically, first  is the nilpotency of order two (i.e. ${s}_{b} ^2 = 0,\; s_{ab}^2 = 0$), and  
second is the absolute anticommutativity (i.e. $s_b\,s_{ab} + s_{ab}\,s_{b} = 0$) of symmetry transformations. 
The first property signifies  that both the  quantum  BRST and  anti-BRST symmetry transformations 
are fermionic in nature whereas  the second property, physically,  shows that both symmetries 
are linearly independent of each other. The key signature of a BRST-quantized non-Abelian gauge theory is the 
existence of the Curci-Ferrari (CF)-condition  at the quantum level [10]. The CF-condition is (anti-)BRST 
invariant quantity which shows that it is a physical condition at the quantum level. 
On the other hand,  the advanced methods of covariant quantization for general gauge theories are based either on
the BRST symmetry principle realized in the well-known quantization scheme [11] by Batalin and
Vilkovisky (BV) or on the extended BRST symmetry principle  realized within the quantization
method [12-14] proposed by Batalin, Lavrov and Tyutin (BLT).

The mass generation in gauge theory is an important aspect [15, 16]. However, it has been observed  that the 4D 
topological massive (non-)Abelian gauge theories have been studied [17-20] within the ambit of BRST formalism  where (non-)Abelian 1-form gauge field 
acquires a mass in a very natural way without taking any help of the Higgs mechanism. 
However, these models have issues with renormalization, consistency and unitarity. 
The BRST analysis of these models has been performed with the hope of reducing these issues [21-26]. 
But, it remains still an open problem to construct a theory that is free from the above issues.

The topology of odd-dimensional spacetime  allows the construction  of gauge theories with novel and attractive features.  
Therefore, in lower dimension, it is interesting to propose an odd-dimensional  massive non-Abelian  1-form gauge invariant  model which is
 free from the above issues (i.e. renormalization, consistency and unitarity).
One such massive non-Abelian 1-form model is the Jackiw-Pi (JP) model, proposed by Jackiw and Pi,  in three (2 + 1)-dimensions (3D) of spacetime
where the parity is respected due to the presence of 1-form vector field [27]. This three  dimensional (3D) model allows the construction 
 of gauge theory with the mass term in a natural fashion.  The JP model  has been studied earlier in the different
prospectives of theoretical interest [27-31].  This model is endowed with  the set of two interesting continuous symmetry transformations
which are  Yang-Mills (YM) and non-Yang-Mills (NYM) symmetry transformations [30, 31].

The usual superfield approach (USFA) to BRST formalism [32-36] exploits the idea
of horizontality condition (HC) for the derivation of off-shell
nilpotent (anti-)BRST symmetry transformations for the gauge, ghost, and anti-ghost fields
where full super expansions of the superfield [with two Grassmannian variables
($\vartheta, \bar\vartheta$)] have been taken into account.  
The USFA does not explain the derivation of the matter fields in an interacting theory. 
This approach is generalized to derive the (anti-)BRST symmetries for the matter fields where 
the idea of HC and gauge invariant restriction(s) (GIR) are used together [37-39].
This extended version of USFA is known as augmented version of the 
superfield approach (AVSA).
The superfield approach has also been extensively discussed  for the general gauge theories 
which provide the geometrical interpretations of the BRST quantization  within the framework of 
 superfield approach to BRST formalism [40-42].  The BRST analysis of the non-Abelian  JP model 
has been performed within the framework of AVSA to BRST formalism [30, 31].
Against the backdrop of the above discussions, we have applied the newly proposed (anti-)chiral 
superfield formalism (ACSA) [43-47] to derive the complete set of (anti-)BRST symmetries as well as CF-condition of the theory where 
(anti-)chiral super expansions of the superfields (i.e. only one Grassmannian variable in expansions) have been taken into account.
The combination of ACSA and modified Bonora-Tonin superfield approach (MBTSA) has been utilized to derive  the 
complete set of (anti-)BRST symmetry transformations for the various reparameterization invariant 
 models  [48, 49].  In our present endeavor, for the first time, the BRST analysis of the odd-dimensional theory 
(i.e. massive 3D non-Abelian JP model) is discussed within the realm  of ACSA to BRST formalism.

In the present paper, the subject matters of  different sections are organized as follows. In Sec. 2, we discuss 
a couple of local gauge symmetry transformations (i.e. YM and NYM symmetries)  for the JP model. Our Sec. 3 deals with the
 coupled (but equivalent) Lagrangian densities and its (anti-)BRST symmetry transformations.
Sec. 4  is fully devoted to the derivation of the conserved (anti-)BRST charges, nilpotency 
and absolute anticommutativity properties of  charges in ordinary spacetime.
Our Sec. 5 contains the explicit derivation of the complete set of (anti-)BRST symmetry transformations
 within the realm of ACSA to BRST formalism.  In Sec. 6, we express the conserved (anti-)BRST charges onto
the (3, 1)-dimensional super sub-manifolds [of the {\it general}  (3, 2)-dimensional supermanifold]
 on which our theory is generalized and provides the proof of nilpotency and absolute
anticommutativity properties of the charges   within the ambit  of ACSA
to BRST formalism. In Sec. 7, we discuss the (anti-)BRST  invariances of the Lagrangian densities 
within the realm of ACSA. Finally, in Sec. 8, we highlight the most important findings 
and conclusions of present endeavor, as well as a few future issues and research prospects.\\

\newpage

\noindent
{\it Convention and Notations}:  We acquire, in the present investigation,  the convention and  notations of
 the spacetime Minkowskian metric as: $ \eta_{\mu\nu} =$ diag$(-1, +1, +1)$, totally 
antisymmetric 3D Levi-Civita  tensor $\varepsilon_{\mu\nu\eta}$ satisfies
$\varepsilon_{\mu\nu\eta}\,\varepsilon^{\mu\nu\eta} = - 3!,\;$ $ 
\varepsilon_{\mu\nu\eta}\,\varepsilon^{\mu\nu\kappa} = - 2! \;\delta^\kappa_\eta,$ etc., 
and $\varepsilon_{012} = + 1 = - \varepsilon^{012}$. The  Greek indices 
$\mu, \nu, \eta,... = 0, 1, 2$  denote the time and space directions. 
We take the dot and cross products as: $L \cdot M = L^a \,M^a, \;L \times M = f^{abc}\, L^a\, M^b\, T^c$
 between a set of two non-null vectors $(L^a,  \,M^a)$ 
in the $SU(N)$ Lie algebraic space where the generators $T^a$ satisfy 
the commutation relation  $[T^a, T^b] = i f^{abc}\, T^c$ with $a, b, c,... = 1, 2, 3,..., N^2 - 1$. The structure 
constants $f^{abc}$ are taken to be totally antisymmetric in $a, b, c$ indices  for 
the $SU(N)$ Lie algebra. We have also acquired   the convention of left-derivative technique 
in all relevant calculations  with respect to the fermionic fields $(\bar C, \, C)$.

\section{Preliminaries: Gauge Symmetries}

We begin with the gauge invariant  Lagrangian density of the three (2 + 1)-dimensional (3D) massive
non-Abelian 1-form gauge theory proposed by Jackiw and Pi [27, 30, 31] 
\begin{eqnarray}
{\cal L} & = & - \frac{1}{4}\; {\cal F}^{\mu\nu} \cdot {\cal F}_{\mu\nu} - \frac{1}{4}\; \big({\cal {G}}^{\mu\nu} 
+ g\; {\cal F}^{\mu\nu} \times \rho\big) \cdot \big({\cal {G}}_{\mu\nu} 
+ g \;{\cal F}_{\mu\nu} \times \rho\big)\nonumber\\
& + & \frac {m}{2}\;\varepsilon^{\mu\nu\eta} \; {\cal F}_{\mu\nu} \cdot \phi_\eta,
\end{eqnarray}
where 2-form [${\cal F}^{(2)} =  \frac {1}{2!} 
(dx^\mu \wedge dx^\nu)\, {\cal F}_{\mu\nu} \cdot T$)] field  strenght  tensor ${\cal F}_{\mu\nu} = \partial_\mu {\cal A}_\nu 
- \partial_\nu {\cal A}_\mu - g \;({\cal A}_\mu \times {\cal A}_\nu)$, corresponding to the 1-form 
$[{\cal A}^{(1)} = dx^\mu {\cal A}_\mu \cdot T]$ vector field ${\cal A}_\mu$, is derived from the 
Maurer-Cartan equation ${\cal F}^{(2)} = d {\cal A}^{(1)} + i\, g \;({\cal A}^{(1)} \wedge {\cal A}^{(1)})$. Similarly, other  field strength tensor 
${\cal {G}}_{\mu\nu} = {\cal D}_\mu \phi_\nu - {\cal D}_\nu \phi_\mu$, corresponding to the 1-form 
$[\phi^{(1)} = dx^\mu\, \phi_\mu \cdot T]$ vector field $\phi_\mu$, is obtained from 
${\cal {G}}^{(2)} = d \phi^{(1)} + i\, g \;[\phi^{(1)} \wedge {\cal A}^{(1)} 
+   {\cal A}^{(1)} \wedge \phi^{(1)}] \equiv \frac {1}{2!} (dx^\mu \wedge dx^\nu)\, {\cal {G}}_{\mu\nu} \cdot T$ where the 
covariant derivative is defined as: ${\cal D}_\mu \phi_\nu = \partial_\mu \phi_\nu - g \;({\cal A}_\mu \times \phi_\nu)$. 
In the above, vector fields  ${\cal A}_\mu$ and $\phi_\mu$ have opposite parity to conserve the parity of the Lagrangian density, 
$\rho$ is  a scalar field (auxiliary field), $m$ is the mass parameter and  $g$ is a coupling constant.

The above Lagrangian density (${\cal L}$) obeys a couple of local and continuous gauge symmetry transformations, namely;
YM gauge transformation $(\delta_g)$ and NYM gauge transformations $(\bar\delta_g)$. 
The infinitesimal version of these symmetries are given as  
\begin{eqnarray}
&& \delta_g \phi_\mu = - g\,(\phi_\mu \times \Sigma), \;\;\qquad\qquad
\delta_g \rho = -\, g\,(\rho \times \Sigma), \quad\qquad \delta_g {\cal A}_\mu = {\cal D}_\mu \Sigma, \nonumber\\
&&\delta_g {\cal F}_{\mu\nu} = - \,g\,({\cal F}_{\mu\nu} \times \Sigma), \;\quad 
\qquad \delta_g {\cal {G}}_{\mu\nu} = - \,g\,({\cal {G}}_{\mu\nu} \times \Sigma),
\end{eqnarray} 
\begin{eqnarray}
&&\bar\delta_g {\cal A}_\mu = 0, \qquad  \qquad \bar\delta_g \phi_\mu = {\cal D}_\mu \Lambda , \qquad  \qquad \bar\delta_g \rho 
= +\; \Lambda,  \nonumber\\
&& \bar\delta_g {\cal F}_{\mu\nu} = 0, \;\;\quad \qquad \bar\delta_g {\cal {G}}_{\mu\nu} 
= - \,g\,({\cal F}_{\mu\nu} \times \Lambda),
\end{eqnarray} 
where $\Sigma = \Sigma \cdot T \equiv \Sigma^a \;T^a$ and $\Lambda   = \Lambda  \cdot T \equiv 
\Lambda ^a\; T^a$ are the infinitesimal  gauge parameters. 
It is straightforward to check that the Lagrangian density (1) transforms under the local and continuous 
 gauge transformations (2) and (3) as 
\begin{eqnarray}
\delta_g {\cal L} = 0, \quad \qquad\qquad \bar\delta_g {\cal L} = \partial_\mu \Big[\frac {m}{2}\; 
\varepsilon^{\mu\nu\eta} \;{\cal F}_{\nu\eta} \cdot \Lambda  \Big].
\end{eqnarray}
Hence, the action integral $S = \int\,d^3\,x\,{\cal L}$ remains invariant under  both the 
YM and NYM gauge transformations (i.e. $ \delta_g \, S = 0, \;  \bar \delta_g \, S = 0)$.
These two gauge transformations $\delta_g$ and $\bar\delta_g$ are independent of each other. Therefore, in the present endeavor,    we shall focus only 
on the BRST analysis of the YM gauge transformations within the ambit of ACSA.

\section{Coupled Lagrangian Densities: Off-Shell  Nilpotent Quantum (Anti-)BRST Symmetries}

In this section, we discuss the construction  of the (anti-)BRST invariant coupled (but equivalent) Lagrangian densities
for the three (2 + 1)-dimensional (3D)  massive non-Abelian 1-form  JP model and its off-shell nilpotent and absolutely  anticommuting (anti-)BRST symmetry
transformations corresponding to the YM gauge transformations. The Lagrangian densities for this model  are;  
\begin{eqnarray}
{\cal L}_{\cal B} &=& - \frac{1}{4}\; {\cal F}^{\mu\nu} \cdot {\cal  F}_{\mu\nu} - \frac{1}{4}\; \big({\cal {G}}^{\mu\nu} 
+ g\, {\cal F}^{\mu\nu} \times \rho\big) \cdot \big({\cal {G}}_{\mu\nu} 
+ g\, {\cal F}_{\mu\nu} \times \rho\big)
+ \frac {m}{2}\;\varepsilon^{\mu\nu\eta} \; {\cal F}_{\mu\nu} \cdot \phi_\eta \nonumber\\
&+&  {\cal B} \cdot (\partial^\mu {\cal A}_\mu) + \frac {1}{2} ({\cal B} \cdot {\cal B} + \bar {\cal B} \cdot \bar {\cal B})
- i\; \partial^\mu \bar C \cdot {\cal D}_\mu C, 
\end{eqnarray}
\begin{eqnarray}
{\cal L}_{\bar {\cal B}} &=& -  \frac{1}{4}\; {\cal F}^{\mu\nu} \cdot {\cal F}_{\mu\nu} - \frac{1}{4}\; \big({\cal {G}}^{\mu\nu} 
+ g\, {\cal F}^{\mu\nu} \times \rho\big) \cdot \big({\cal {G}}_{\mu\nu} 
+ g\, {\cal F}_{\mu\nu} \times \rho\big) 
+ \frac {m}{2}\;\varepsilon^{\mu\nu\eta} \; {\cal F}_{\mu\nu} \cdot \phi_\eta \nonumber\\
&-& \bar {\cal B} \cdot (\partial^\mu {\cal A}_\mu) + \frac {1}{2} ({\cal B} \cdot {\cal B} + \bar {\cal B} \cdot \bar {\cal B}) 
- i\; {\cal D}^\mu \bar C \cdot \partial_\mu C,
\end{eqnarray}
where ${\cal B}$ and $\bar {\cal B}$  are the Nakanishi-Lautrup type 
auxiliary fields that have been introduced  for linearizing the gauge fixing terms
which are connected with the Curci-Ferrari condition:
 ${{\cal B} + \bar {\cal B} + i\,g\,(\bar C\times C)} = 0 $ where the 
(anti-)ghost fields ${\bar C}$ and $C$ are fermionic
  [i.e. $(C^a)^2= ({\bar C^a}) ^2=0, \;  C^a C^b+C^b C^a= 0,\; C^a \bar C^b+ \bar C^b C^a = 0,\; \bar C^a\bar C^b+\bar 
C^b\bar C^a = 0,\; \bar C^a C^b+C^b\bar C^a= 0,$ etc.] in nature.
In the above, we have the covariant derivatives 
$[{\cal D}_{\mu} C = \partial_{\mu} C - g\, ({\cal A}_{\mu}\times C) $ 
and ${\cal D}_{\mu}\bar C = \partial_{\mu}\bar C - g\, ({\cal A}_{\mu}\times\bar C)]$   in 
the adjoint representation.

The complete set of off-shell nilpotent and absolutely anticommuting (anti-)BRST symmetry transformations 
corresponding to  the above coupled (but equivalent) Lagrangian  densities [Eqs. (5) and (6)] are as follows:
\begin{eqnarray}
&&s_{ab} {\cal A}_\mu = {\cal D}_\mu \bar C, \;\quad s_{ab} \bar C = \frac {g}{2} \,(\bar C \times \bar C), 
 \quad \; s_{ab} {\cal  B} = - g\,({\cal  B} \times \bar C),\nonumber\\ 
&& s_{ab} \rho = - \,g\,(\rho \times \bar C),\qquad  s_{ab} \phi_\mu = - \,g\,(\phi_\mu \times \bar C),\qquad 
s_{ab} C = i \,\bar {\cal B},\nonumber\\
&& s_{ab} \bar {\cal  B} = 0, \quad s_{ab} {\cal F}_{\mu\nu} = - \,g\,({\cal F}_{\mu\nu} \times \bar C),
\quad s_{ab} {\cal G}_{\mu\nu} = - \,g\,({\cal G}_{\mu\nu} \times \bar C), 
\end{eqnarray}
\begin{eqnarray}
&& s_b {\cal A}_\mu = {\cal D}_\mu C, \qquad s_b C =  \frac{g}{2}\; (C \times C), 
\qquad s_b \bar {\cal  B} = -\, g\,(\bar {\cal  B} \times C),  \nonumber\\
&& s_b \rho = - \,g\, (\rho \times C), \qquad s_b \phi_\mu = - \,g\,(\phi_\mu \times C), \qquad s_b \bar C = i\, {\cal  B},\nonumber\\
&& s_b {\cal  B} = 0, \quad  s_b {\cal  F}_{\mu\nu} = -\, g\,({\cal  F}_{\mu\nu} \times C), \qquad  s_b {\cal  G}_{\mu\nu} = -\, g\,({\cal  G}_{\mu\nu} \times C).
\end{eqnarray}  
The Lagrangian densities [Eqs. (5) and (6)] can be written in term of the starting Lagrangian density (1) plus 
(anti-)BRST symmetry transformations of the some specific terms as;
\begin{eqnarray}
{\cal L}_{\cal B} = {\cal L}_0 + s_b \, s_{ab} \bigg[\frac {i}{2} \, {\cal A}_\mu \cdot {\cal A}^\mu - \bar C \cdot  C 
+ \frac {1}{2}\, \phi_\mu \cdot \phi^\mu \bigg],\nonumber\\
{\cal L}_{\bar {\cal B}} = {\cal L}_0 - s_{ab} \, s_b \bigg[\frac {i}{2} \, {\cal A}_\mu \cdot {\cal A}^\mu 
- \bar C \cdot C + \frac {1}{2}\, \phi_\mu \cdot \phi^\mu \bigg],
\end{eqnarray}
above expressions suggest  the idea of construction of the Lagrangian densities [Eqs. (5) and (6)] 
where each term of the square brackets are chosen such that all the individual terms are Lorentz scalar, 
 having equal ghost number (i.e. zero)  and  equal mass dimension (i.e. one).

The anticommutators of the (anti-)BRST symmetries $[s_{(a)b}]$ for the various fields and field strength tensors present in the
Lagrangian densities [Eqs. (5) and (6)] are  given as:
\begin{eqnarray}
&& \{s_b,\; s_{ab}\} \;{\cal A}_\mu = \; i\, \partial_\mu \, [{\cal B} + \bar {\cal B} + i\, g\, (\bar C \times C)] + i\,g\, 
[{\cal B} + \bar {\cal B} + i\, g\, (\bar C \times C)]\times {\cal A}_\mu ,  \nonumber\\
&& ~~~~~~~~~~~~~~~~ = \;i\,  {\cal D}_\mu [{\cal B} + \bar {\cal B} + i\, g\, (\bar C \times C)], \nonumber\\
&& \{s_b,\; s_{ab}\} \;\phi_\mu \; = \;  i\, g\, [{\cal B} + \bar {\cal B} + i\, g\, (\bar C \times C)]\times \phi_\mu, \nonumber\\
&& \{s_b,\; s_{ab}\} \;\rho \;\;\; = \; i\,g\, [{\cal B} + \bar {\cal B} + i\, g\, (\bar C \times C)]\times \rho, \nonumber\\
&& \{s_b,\; s_{ab}\} \;{\cal F}_{\mu\nu} = \; i\,g\, [{\cal B} + \bar {\cal B} + i\, g\, (\bar C \times C)] \times {\cal F}_{\mu\nu}, \nonumber\\
&& \{s_b,\; s_{ab}\} \;{\cal {G}}_{\mu\nu} \,= \; i\,g\, [{\cal B} + \bar {\cal B} + i\, g\, (\bar C \times C)] \times {\cal {G}}_{\mu\nu}.  
\end{eqnarray}
Hence, it is clear that absolute anticommutativity of (anti-)BRST symmetries for various fields  is satisfied 
 if and only if CF-condition
[${\cal B} + \bar {\cal B} + i\, g\, (\bar C \times C) = 0$] is satisfied. 
The anticommutators for the rest of the fields directly come out to be zero. Thus, ultimately,  we have the following relationships   
\begin{eqnarray}
&& \{s_b,\; s_{ab}\} \;{\cal A}_\mu =  0, \qquad \; \{s_b,\; s_{ab}\} \;\phi_\mu =  0, \qquad \;\;\,\{s_b,\; s_{ab}\} \;\rho = 0, \nonumber\\
&& \{s_b,\; s_{ab}\} \;{\cal F}_{\mu\nu} = 0, \qquad \{s_b,\; s_{ab}\} \;{\cal {G}}_{\mu\nu} = 0, \qquad \; \{s_b,\; s_{ab}\} \; C =  0,\nonumber\\
&& \{s_b,\; s_{ab}\} \; \bar C =  0, \qquad\;\;\; \{s_b,\; s_{ab}\} \;{\cal B} =  0, \qquad \;\;\;\; \{s_b,\; s_{ab}\} \;{\cal {\bar B}} =  0. 
\end{eqnarray}

It can be checked that the preceding  Lagrangian densities (${\cal L}_{\cal B}$ and ${\cal L}_{\bar {\cal B}}$) transform 
under the off-shell nilpotent (anti-)BRST transformations, as
\begin{eqnarray}
s_b {\cal L}_{\cal B} &=& \; \partial_\mu[{\cal B}\cdot ({\cal D}^\mu C)], \quad \qquad
s_{ab} {\cal L}_{\bar {\cal B}} = - \,\partial_\mu[\bar {\cal B}\cdot ({\cal D}^\mu \bar C)],\nonumber\\
s_{ab} {\cal L}_{\cal B} &=& \; \partial_\mu({\cal B} \cdot \partial^\mu \bar C) 
- {\cal D}_\mu [{\cal B} + \bar {\cal B} + i\, g\, (\bar C \times  C)] \cdot \partial^\mu \bar C, \nonumber\\
s_b {\cal L}_{\bar {\cal B}} &= &-\, \partial_\mu(\bar {\cal B} \cdot \partial^\mu C) 
+ {\cal D}_\mu [{\cal B} + \bar {\cal B} + i\, g  (\bar C \times C)] \cdot \partial^\mu C.
\end{eqnarray}
Thus, the corresponding actions (i.e. $S_{\cal B} = \int d^2 x \;{\cal L}_{\cal B}$ 
and $S_{\bar {\cal B}} = \int d^2 x \;{\cal L}_{\bar {\cal B}}$) remain invariant under the (anti-)BRST
symmetries  in the  three (2 + 1)-dimensional  (3D) ordinary spacetime manifold where the CF-condition is satisfied. \\

\section{(Anti-)BRST  Currents and Charges: Nilpotency and Absolute Anticommutativity Properties}

In this section, we discuss  the conserved currents and conserved charges  corresponding to the 
(anti-)BRST symmetry transformations using  the Noether theorem\footnote{According to the Noether theorem
 whenever any Lagrangian density or its corresponding action remains invariant under any  
continuous symmetry transformations, there exits conserved currents and charges corresponding
to that given continuous symmetries.}.  We prove  the nilpotency and absolute anticommutativity 
properties of the conserved charges within the framework of BRST formalism. Towards this aim in mind, 
first of all, we derive  the Noether (anti-)BRST currents ${\cal J}^\mu_{(a)b}$ for the 3D JP model as:   
\begin{eqnarray} 
{\cal J}^\mu_b &=& {\cal B} \cdot ({\cal D}^\mu C) - \big[{\cal F}^{\mu\nu} - g\,(({\cal {G}}^{\mu\nu} 
+ g \,{\cal F}^{\mu\nu} \times \rho) \times \rho) 
- m \,\varepsilon^{\mu\nu\eta} \, \phi_\eta \big]  \cdot ({\cal D}_\nu C)\nonumber\\
&+& g \,[{\cal {G}}^{\mu\nu} + g\, ({\cal F}^{\mu\nu} \times \rho) ]\cdot  (\phi_\nu \times C ) 
+ \frac {i}{2} \; g \; \partial^\mu \bar C \cdot (C \times C),
\end{eqnarray}
\begin{eqnarray} 
{\cal J}^\mu_{ab} &=& - \bar {\cal B} \cdot ({\cal D}^\mu \bar C) - \big[{\cal F}^{\mu\nu} - g\,(({\cal {G}}^{\mu\nu} 
+ g\, {\cal F}^{\mu\nu} \times \rho) \times \rho) 
- m \,\varepsilon^{\mu\nu\eta} \, \phi_\eta \big]  \cdot ({\cal D}_\nu \bar C)\nonumber\\ 
&+& g\, [{\cal {G}}^{\mu\nu} + g (F^{\mu\nu} \times \rho) ]\cdot  (\phi_\nu \times \bar C ) 
- \frac {i}{2} \;g \; \partial^\mu C \cdot (\bar C \times \bar C).
\end{eqnarray}
These expressions for  the Norther (anti-)BRST currents can be re-expressed for the algebraic convenience in the following form:  
\begin{eqnarray}
{\cal J}^\mu_b &\;=\;& {\cal B}  \cdot ({\cal D}^\mu C) - \partial^\mu {\cal B} \cdot C 
- \frac {i}{2} \;g \; \partial^\mu \bar C \cdot (C \times C)\nonumber\\
&\;-\;& \partial_\nu [({\cal F}^{\mu\nu} - g\, \{({\cal {G}}^{\mu\nu} 
+ g\, {\cal F}^{\mu\nu} \times \rho) \times \rho\} 
- m \;\varepsilon^{\mu\nu\eta} \; \phi_\eta)\cdot C],
\end{eqnarray}
\begin{eqnarray}
{\cal J}^\mu_{ab} &\;=\;&  - \bar {\cal B}  \cdot ({\cal D}^\mu \bar C) + \partial^\mu \bar {\cal B}  \cdot {\bar C} 
+ \frac {i}{2} \;g \; \partial^\mu C \cdot (\bar C \times \bar C)\nonumber\\
&\,\,-\;& \partial_\nu [({\cal F}^{\mu\nu} - g\, \{({\cal {G}}^{\mu\nu} + g\, {\cal F}^{\mu\nu} \times \rho) \times \rho\} 
- m \;\varepsilon^{\mu\nu\eta} \; \phi_\eta)\cdot \bar C].
\end{eqnarray}
The conservation law [i.e. $\partial_\mu {\cal J}^\mu_{(a)b} = 0$] of the above (anti-)BRST currents  can be 
proven  by exploiting the Euler-Lagrange equations of motion (EL-EoMs) derived from the 
 Lagrangian densities ${\cal L}_{\cal B}$ and ${\cal L}_{\bar {\cal B}}$, respectively, as:
\begin{eqnarray}
&&{\cal D}_\mu {\cal F}^{\mu\nu} - g\, {\cal D}_\mu [({\cal {G}}^{\mu\nu} + g\, {\cal F}^{\mu\nu}\times \rho) \times \rho] 
+ m \, \varepsilon^{\mu\eta\nu} \, {\cal D}_\mu \phi_\eta - \partial^\nu {\cal B}  \nonumber\\
&&+ g \,[({\cal {G}}^{\mu\nu} + g\, {\cal F}^{\mu\nu}\times \rho) \times \phi_\mu]  
- i \, g\, (\partial^\nu \bar C \times C) = 0, \nonumber\\
&&{\cal D}_\mu [{\cal {G}}^{\mu\nu} + g \,({\cal F}^{\mu\nu}\times \rho)] 
+ \frac {m}{2} \, \varepsilon^{\mu\eta\nu} \, {\cal F}_{\mu\eta} = 0, \nonumber\\
&&({\cal {G}}^{\mu\nu} + g\, {\cal F}^{\mu\nu}\times \rho) \times {\cal F}_{\mu\nu} = 0, \quad \quad \partial_\mu ({\cal D}^\mu C) = 0,  
\quad \quad {\cal D}_\mu (\partial^\mu \bar C) = 0,
\end{eqnarray}
\begin{eqnarray}
&&{\cal D}_\mu {\cal F}^{\mu\nu} - g\, {\cal D}_\mu [({\cal {G}}^{\mu\nu} + g\, {\cal F}^{\mu\nu}\times \rho) \times \rho] 
+ m \, \varepsilon^{\mu\eta\nu} \, {\cal D}_\mu \phi_\eta + \partial^\nu \bar {\cal B}  \nonumber\\
&&  + g \,[({\cal {G}}^{\mu\nu} + g\, {\cal F}^{\mu\nu}\times \rho) \times \phi_\mu]  
+ i \, g\, ( \bar C \times \partial^\nu C) = 0, \nonumber\\
&& {\cal D}_\mu [{\cal {G}}^{\mu\nu} + g \,({\cal F}^{\mu\nu}\times \rho) ] 
+ \frac {m}{2} \, \varepsilon^{\mu\eta\nu} \; {\cal F}_{\mu\eta} = 0, \nonumber\\
&& ({\cal {G}}^{\mu\nu} + g\, {\cal F}^{\mu\nu}\times \rho) \times {\cal F}_{\mu\nu} = 0, \quad
\quad \partial_\mu ({\cal D}^\mu \bar C) = 0,  \quad\quad {\cal D}_\mu (\partial^\mu C) = 0.
\end{eqnarray}
Now the expressions of the Noether conserved currents  lead to the derivation 
of the following conserved (anti-)BRST charges using  ${\cal Q}_{(a)b} = \int d^2 x \; {\cal J}^0_{(a)b}$:
\begin{eqnarray}
&&{\cal Q}_{ab} = -  \int d^2 x  \Big[\bar {\cal B} \cdot {\cal D}^0 \bar C 
- {\dot {\bar {\cal B}}} \cdot \bar C 
- \frac{i}{2}\, g\,\dot C \cdot (\bar C \times \bar C) \Big],
\end{eqnarray}
\begin{eqnarray}
&&{\cal Q}_b = \int d^2 x  \Big[{\cal B} \cdot {\cal D}^0 C - \dot {\cal B} \cdot C 
- \frac{i}{2}\,g\, \dot {\bar C} \cdot (C \times C) \Big].
\end{eqnarray}
It is straightforward to check that the above conserved charges [${\cal Q}_{(a)b}$] are nilpotent 
 of order two (i.e. ${\cal Q}_b^2 = {\cal Q}_{ab}^2 = 0)$ and they obey absolute anticommutativity property 
(i.e. $Q_b \, {\cal Q}_{ab} + {\cal Q}_{ab}\,{\cal Q}_b = 0$) in the ordinary 
(2 + 1)-dimensional (3D) spacetime. These two properties are captured by the definition of generator 
 expressed as follows:
\begin{eqnarray}
&&s_b {\cal Q}_b \;\;\,\,  = -i\,\,{\{{\cal Q}_b,{\cal Q}_b}\} = 0 \;\;\, \Longrightarrow \;\; \, {\cal Q}_b^2 = 0,\nonumber\\
&&s_{ab} \,{\cal Q}_{ab} = -i\,\{{\cal Q}_{ab},{\cal Q}_{ab}\} = 0 \,\Longrightarrow \,\;\;  {\cal Q}_{ab}^2 = 0,\nonumber\\
&&s_{ab}{\cal Q}_{b} \;\; = -i\,{\{{\cal Q}_{b},{\cal Q}_{ab}}\} = 0\; \; \Longrightarrow \;\; \{{\cal Q}_{b},{\cal Q}_{ab}\} = 0, \nonumber\\ 
&& s_b {\cal Q}_{ab}  \;\; = -i\,{\{{\cal Q}_{ab},{\cal Q}_b}\} = 0 \; \; \Longrightarrow \;\; \{{\cal Q}_{ab},{\cal Q}_{b}\} = 0. 
\end{eqnarray} 
It is evident that, from the above expressions,  the  nilpotency property is satisfied on the direct application of (anti-)BRST symmetry transformations 
 on the conserved charges [Eqs. (19) and (20)] whereas absolute anticommutativity property
is satisfied if and only if CF-condition [${\cal B} + \bar {\cal B} + i\,g\, (\bar C \times C) = 0$]
is satisfied\footnote{We obtain  the following expressions  after the application of $s_b$ on ${\cal Q}_{ab}$
and $s_{ab}$ on ${\cal Q}_{b}$ for the proof of absolute anticommutativity as: $s_b {\cal Q}_{ab} = -\, i\int d^2x 
\Big[\bar {\cal B} \cdot \partial^0 \Big({\cal B} + \bar {\cal B} 
\,+ \,i \, g\, (\bar C\times  C)\Big)\Big] = -\,i\,\{{\cal Q}_b, {\cal Q}_{ab}\} = 0$ and 
$s_{ab} {\cal Q}_b =  i \int d^2x \Big[{\cal B} \cdot \partial^0 \Big({\cal B} + \bar {\cal B} 
\,+ \,i \, g\, (\bar C\times C)\Big)\Big] = -\,i\,\{{\cal Q}_{ab}, {\cal Q}_{b}\} = 0$.}.

The (anti-)BRST conserved charges cab be written in terms of the (anti-)BRST exact  form 
with respect to the (anti-)BRST symmetries $(s_{(a)b})$ as: 
\begin{eqnarray}
&&{\cal Q}_b = \int d^2 x\; s_b \; \Big[{\cal B} (x) \cdot {\cal A}_0 (x) + i \; \dot{\bar C} (x) \cdot C(x) \Big], \nonumber\\
&&{\cal Q}_{ab} = \int d^2 x \; s_{ab}\;  \Big [ \bar {\cal B} (x) \cdot {\cal A}_0(x) + i \; \dot C (x) \cdot \bar C (x) \Big], 
\end{eqnarray}
\begin{eqnarray}
&&{\cal Q}_b = s_{ab}\;\int d^{2}x\;\Big[i\,C\cdot\dot C - \frac {1}{2}\; C\cdot ({\cal A}_0 \times C)\Big], \nonumber\\
&&{\cal Q}_{ab} = s_b \;\int d^{2}x\;\Big[-\;i\;{\bar C}\cdot \dot{\bar C} + \frac {1}{2}\; {\bar C} \cdot ({\cal A}_0\times \bar C)\Big].
\end{eqnarray}
It is crystal clear to prove the nilpotency and absolute 
anticommutativity  properties for  the conserved (anti-)BRST charges [${\cal Q}_{(a)b}$] in 
quite straightforward manner using the above (anti-)BRST exact form [Eqs. (22) and (23)] which  demonstrate the alternate 
proof of these two properties for the (anti-)BRST charges.

\section {Off-Shell Nilpotent (Anti-)BRST Symmetry Transformations: (Anti-)chiral Superfield Approach}

In this section,  we derive the off-shell nilpotent (anti-)BRST symmetry transformations by exploiting the (anti-)chiral superfields 
approach (ACSA) to BRST formalism where we use the (anti-)chiral super expansions of the (anti-)chiral  superfields. Towards this aim in mind,
first of all, we generalize the ordinary fields of Lagrangian densities [Eq. (5) and (6)] (of three (2 + 1)-dimensional (3D) spacetime)  onto the 
the (3, 1)-dimensional (anti-)chiral super sub-manifold of the suitably chosen general (3, 2)-dimensional supermanifold.\\

\subsection {BRST Symmetry Transformations: ACSA} \vskip 0.2 cm 

In this subsection,  we concentrate on the derivation of the BRST symmetry transformations
 for all the fields of Lagrangian densities [Eq. (5) and (6)] using the anti-chiral super 
expansions of the superfields. For this, we generalize three (2 + 1)-dimensional  basic and auxiliary 
ordinary fields onto (3, 1)-dimensional super sub-manifold  as   
\begin{eqnarray}
&& {\cal A}_\mu (x) \longrightarrow \; B_\mu (x, \bar\vartheta) = {\cal A}_\mu (x) + \bar\vartheta\, {\cal R}_\mu (x),\nonumber\\
&& \phi_\mu (x) \; \longrightarrow \; \tilde\Phi_\mu (x, \bar\vartheta) = \phi_\mu (x) + \bar\vartheta\, {\cal S}_\mu (x),\nonumber\\
&& C (x) \;\;\longrightarrow  \; {\cal F} (x, \bar\vartheta) = C(x) + i\,\bar\vartheta\, B_1 (x),\nonumber\\
&& \bar C (x) \;\;\longrightarrow  \; \bar {\cal F} (x, \bar\vartheta) = \bar C(x) + i\,\bar\vartheta\,  B_2 (x),\nonumber\\
&& {\cal B} (x) \;\;\longrightarrow \;\tilde {\cal B} (x, \bar\vartheta) = {\cal B} (x) + \bar\vartheta\, f_1 (x),\nonumber\\
&& \bar {\cal B} (x) \;\;\longrightarrow \; {\tilde {\bar {\cal B}}} (x, \bar\vartheta) = \bar {\cal B} (x) + \bar\vartheta\, f_2 (x),\nonumber\\
&& \rho (x) \;\;\;\longrightarrow \; {\tilde \rho} (x, \bar\vartheta) = \rho (x) + \bar\vartheta\, f_3 (x),
\end{eqnarray}
where the coefficients of $\bar\vartheta$  (i.e. ${\cal R}_\mu, \, {\cal S}_\mu, \, B_1, \, B_2, \, f_1,\, f_2,\; f_3$) 
are the secondary fields which have to  determine  by using the key ideas of ACSA. The fermionic nature of  $\bar\vartheta$  ensures that 
 $(B_1, \, B_2)$ are bosonic in nature and $({\cal R}_\mu, \, {\cal S}_\mu, \, f_1,\, f_2,\; f_3$) are fermionic in nature.

Towards this aim of determining the values of secondary fields, we note very usefully  and interesting 
BRST invariant quantities\footnote{The (anti-)BRST invariant quantities are obtained using nilpotency property 
of (anti-)BRST symmetry transformations [Eqs. (7) and (8)] and some of them are determined by the hit and trial method.} 
which are specific combinations of the basic and auxiliary fields of the Lagrangians densities, namely;
\begin{eqnarray}
&& s_b {\cal B} = 0, \quad s_b ({\cal D}_\mu C) = 0, \quad s_b (C\times C) = 0, \quad s_b ({\cal A}_\mu\cdot \partial_\mu {\cal B} 
 + i\, \partial _\mu \bar C\cdot {\cal D}^\mu C) = 0, \nonumber\\
&& s_b (\bar {\cal B}\times C) = 0, \quad s_b (\rho \times C) = 0, \quad s_b (\phi_\mu \times C) = 0, \quad s_{b} ({\cal B}\times \bar C) = 0. 
\end{eqnarray}
According  to the basic principle of the ACSA to BRST formalism, all the above  BRST invariant restrictions 
must be independent of the Grassmannian coordinate $\bar\vartheta$ when these BRST invariant restrictions 
are generalized onto the (3, 1)-dimensional anti-chiral super sub-manifold of the most general (3, 2)-dimensional supermanifold  as
\begin{eqnarray*}
&&\tilde {\cal B} (x, \bar\vartheta) = {\cal B} (x),  \quad 
\partial_\mu {\cal F} (x, \bar\vartheta) -  g\,  B_\mu (x, \bar\vartheta) \times {\cal F} (x, \bar\vartheta)  = \partial_\mu C - 
g\,{\cal A}_\mu (x) \times C(x), \nonumber\\
\end{eqnarray*}
\begin{eqnarray}
&&{\cal F} (x, \bar\vartheta) \times {\cal F} (x, \bar\vartheta) = C(x)\times C(x), \quad B_\mu (x, \bar\vartheta)\cdot
\partial_\mu {\cal B}(x, \bar\vartheta) + i\, \partial _\mu \bar {\cal F} (x, \bar\vartheta) \cdot [\partial^\mu {\cal F}(x, \bar\vartheta)\nonumber\\
&&  - \, g\, B^\mu (x, \bar\vartheta)\times {\cal F}(x, \bar\vartheta)] = {\cal A}_\mu (x)\cdot \partial_\mu {\cal B} (x) + i\, \partial_\mu \bar C (x)\cdot 
[\partial^\mu C(x) - g\, ({\cal A}^\mu (x)\times C(x))], \nonumber\\
&& {\tilde{\bar{\cal B}}} (x, \bar\vartheta) \times {\cal F} (x, \bar\vartheta) = {\bar {\cal B}} (x) \times C(x), \quad 
\tilde \rho (x, \bar\vartheta)\times  {\cal F} (x, \bar\vartheta) = \rho (x) \times C(x),\nonumber\\
 && \tilde \Phi_\mu (x, \bar\vartheta)\times  {\cal F} (x, \bar\vartheta) = \phi_\mu (x) \times C(x), \quad
{\tilde {{\cal B}}} (x, \bar\vartheta)\times  \bar {\cal F} (x, \bar\vartheta) = {\cal B} \times \bar C(x).   
\end{eqnarray}
From  the above generalized quantities, we are able to find out the values of secondary fields present in the 
expansions of the anti-chiral superfields (24) as 
\begin{eqnarray}
&& s_b {\cal B} = 0 \;\Longrightarrow \;\tilde {\cal B} (x, \bar\vartheta) = {\cal B} (x) \Longrightarrow \; \;  f_1 (x) = 0, \nonumber\\
&& s_b (C\times C) = 0 \;\Longrightarrow \; {\cal F} (x, \bar\vartheta) \times {\cal F} (x, \bar\vartheta) = C(x)\times C(x) \Longrightarrow B_1 \times C = 0.
\end{eqnarray}
The latter condition $B_1 \times C = 0$ implies that one of the possible solution is $B_1 \propto (C\times C)$. 
Thus, we have $B_1  = \kappa\, (C\times C)$ where $\kappa$ is proportionality constant which implies 
the modified  form of anti-chiral superfield ${\cal F} (x, \bar\vartheta)$ as
\begin{eqnarray}
&& {\cal F} (x, \bar\vartheta) \longrightarrow {\cal F}^{(m)} (x, \bar\vartheta) = C(x)  + i\,\bar\vartheta\, \kappa\, (C\times C),  
\end{eqnarray}
where superscript $(m)$ denotes the modified form of anti-chiral superfield. Now, we focus on the 
generalization of the $s_b ({\cal D}_\mu C) = 0$, we have the following 
\begin{eqnarray}
\partial_\mu {\cal F}^{(m)} (x, \bar\vartheta) -  g\,  B_\mu (x, \bar\vartheta) \times {\cal F}^{(m)} (x, \bar\vartheta)  & = & \partial_\mu C - 
g\, {\cal A}_\mu (x) \times C(x)\nonumber\\
 \Longrightarrow  {\cal R}_\mu  & = &  \frac {2\,i\,\kappa}{g}\, {\cal D}_\mu C(x).
\end{eqnarray}
As a consequence, we have the modified form of the superfield $B_\mu (x, \bar\vartheta)$ as 
\begin{eqnarray}
B_\mu (x, \bar\vartheta) \longrightarrow B_\mu ^{(m)} (x, \bar\vartheta) = {\cal A}_\mu (x) + \frac {2\,i\,\kappa}{g}\,\bar\vartheta\, {\cal D_\mu} C(x).
\end{eqnarray} 
Now, we use BRST invariant quantity $s_b ({\cal B}\times \bar C) = 0$ which implies the following  
\begin{eqnarray}
{\tilde{{\cal B}}} (x, \bar\vartheta) \times \bar {\cal F} (x, \bar\vartheta) = {{\cal B}} (x) \times \bar C(x) \Longrightarrow B_2 \propto {\cal B}.
\end{eqnarray} 
The above relationship  (i. e. $B_2 \propto {\cal B}$) leads to the value of $ B_2$ as: $B_2 =  {\cal B}$ where numerical constant 
is taken to be unit for the shake of simplicity.

In order to determine the value of constant $\kappa$, we use now the generalization of the BRST invariant 
quantity $s_b ({\cal A}_\mu\cdot \partial_\mu {\cal B} + i\, \partial _\mu \bar C\cdot {\cal D}^\mu C) = 0$ as 
\begin{eqnarray}
B_\mu (x, \bar\vartheta)\cdot
\partial_\mu {\cal B}(x, \bar\vartheta) + i\, \partial _\mu \bar {\cal F} (x, \bar\vartheta) \cdot [\partial^\mu {\cal F}(x, \bar\vartheta) -  g\,(B^\mu (x, \bar\vartheta)\times {\cal F}(x, \bar\vartheta))] \nonumber\\
= {\cal A}_\mu (x)\cdot \partial_\mu {\cal B} (x) + i\, \partial_\mu \bar C (x)\cdot 
[\partial^\mu C(x) - g\, ({\cal A}^\mu (x)\times C(x))],
\end{eqnarray}
after substitutions of the modified anti-chiral super expansions from (28) and (30) into (32), we get $\kappa = -\,  {i\,g}/{2}$.
Thus, finally, we have the values of secondary fields as: 
\begin{eqnarray}
{\cal R}_\mu = {\cal D_\mu} C, \qquad  B_1 = -\, \frac {i\,g}{2} \,(C\times C),\qquad B_2 =  {\cal B}. 
\end{eqnarray}
We now focus on the generalization of the $s_b (\bar {\cal B}\times C) = 0, \;  s_b (\phi_\mu \times C) = 0 $ and 
$s_b (\bar {\cal B}\times \rho) = 0$ which lead to the derivation of secondary fields as:
\begin{eqnarray}
&& {\tilde{\bar{\cal B}}} (x, \bar\vartheta) \times {\cal F} (x, \bar\vartheta) = {\bar {\cal B}} (x) \times C(x) \quad \; \, \Longrightarrow \; 
 f_2 = -\, g\, (\bar {\cal B} \times C),\nonumber\\
&& \tilde \Phi_\mu (x, \bar\vartheta)\times  {\cal F} (x, \bar\vartheta) = \phi_\mu (x) \times C(x) \;\;\Longrightarrow \; {\cal S}_\mu  =
-\,g\, (\phi_\mu \times C), \nonumber \\
&& \tilde \rho (x, \bar\vartheta)\times  {\cal F} (x, \bar\vartheta) = \rho (x) \times C(x) \;\;\;\;\;\;\Longrightarrow \; f_3 =
-\,g\, (\rho \times C).  
\end{eqnarray}
The substitution of the values of secondary fields  from (33) and (34) into the expansions
of the superfields (24) lead to the following 
\begin{eqnarray}
&& {\cal A}_\mu (x) \longrightarrow \; B_\mu ^{(b)} (x, \bar\vartheta)  
= {\cal A}_\mu (x) + \bar\vartheta\, [{\cal D_\mu} C (x)] = 
{\cal A}_\mu (x) + \bar\vartheta\, [s_b {\cal A}_\mu (x)],\nonumber\\
&& \phi_\mu (x) \;\longrightarrow \; \tilde\Phi_\mu ^{(b)} (x, \bar\vartheta) 
= \phi_\mu (x) + \bar\vartheta\, [-\,g\, (\phi_\mu (x) \times C (x))]
= \phi_\mu (x) + \bar\vartheta\, [s_b\phi_\mu (x)],\nonumber\\
&& C (x) \;\;\longrightarrow \;  {\cal F} ^{(b)} (x, \bar\vartheta) 
= C(x) + \bar\vartheta\, [\frac {g}{2} \,(C(x)\times C(x))]
= C(x) + \bar\vartheta\, [s_b C(x)],  \nonumber\\
&& \bar C (x) \;\;\longrightarrow  \; \bar {\cal F} ^{(b)} (x, \bar\vartheta) 
= \bar C(x) + \bar\vartheta\, [i\,{\cal B}(x)]
= \bar C(x) + \bar\vartheta\, [s_b \bar C (x)],\nonumber\\
&& {\cal B} (x) \;\;\longrightarrow \; \tilde {\cal B} ^{(b)} (x, \bar\vartheta) 
\; = {\cal B} (x) + \bar\vartheta\, [0]
= {\cal B} (x) + \bar\vartheta\, [s_b {\cal B} (x)],\nonumber\\
&& \bar {\cal B} (x) \;\;\longrightarrow \; {\tilde {\bar {\cal B}}} ^{(b)} (x, \bar\vartheta) 
\; = \bar {\cal B} (x) + \bar\vartheta\, [-\, g\, (\bar {\cal B}(x) \times C (x))]
= \bar {\cal B} (x) + \bar\vartheta\, [s_b \bar {\cal B} (x)],\nonumber\\
&& \rho (x) \;\;\,\longrightarrow \; {\tilde \rho} ^{(b)} (x, \bar\vartheta) 
\;\;  = \rho (x) + \bar\vartheta\, [-\,g\, (\rho (x) \times C (x))]
= \rho (x) + \bar\vartheta\, [s_b \rho (x)],
\end{eqnarray}
where the superscript $(b)$ on the anti-chiral superfields denotes the anti-chiral superfields that have been obtained after the 
application of the BRST invariant restrictions (25). Here, the coefficients of the $\bar\vartheta$ are nothing but the BRST symmetry 
transformations for various fields [25-29]. Thus, we have derived all the BRST symmetry transformations for coupled (but equivalent) 
Lagrangian densities and shown the sanctity of BRST symmetry transformations  within the ambit of ACSA.
We also conclude that there is a  connection between BRST symmetry ($s_b$) and the translational generator 
$(\partial_{\bar\vartheta})$ along the $\bar\vartheta$-direction of the (3, 1)-dimensional anti-chiral super sub-manifold
 with the following relationship and mapping:
\begin{eqnarray}
\partial_{\bar\vartheta}\, \Omega^{(b)} \,(x, \bar\vartheta) = s_b \, \omega (x), \qquad  s_b\longleftrightarrow \partial_{\bar\vartheta}. 
\end{eqnarray}
 Here it is clear that the BRST symmetry transformations for any ordinary generic field $\omega (x)$
are nothing but the translation of the generalized anti-chiral generic superfield $[\Omega^{(b)} (x, \bar\vartheta)]$ onto the 
(3, 1)-dimensional super sub-manifold along the $\bar\vartheta$-direction.\\

\subsection {Anti-BRST Symmetry Transformations: ACSA} \vskip 0.2 cm

In this subsection, we derive all the anti-BRST symmetry transformations for various fields of the Lagrangians densities 
within the ambit of ACSA to BRST formalism. Towards this aim in mind, first of all, we generalize our basic and auxiliary 
fields onto (3, 1)-dimensional chiral super sub-manifold of the most general (3, 2)-dimensional supermanifold as;  
\begin{eqnarray*}
&& {\cal A}_\mu (x) \longrightarrow B_\mu (x, \vartheta) = {\cal A}_\mu (x) + \vartheta\, \bar {\cal R}_\mu (x),\nonumber\\
&& \phi_\mu (x) \,\longrightarrow \tilde\Phi_\mu (x, \vartheta) = \phi_\mu (x) + \vartheta\, \bar {\cal S}_\mu (x),\nonumber\\
&& C (x) \; \longrightarrow  {\cal F} (x, \vartheta) = C(x) + i\,\vartheta\, \bar B_1 (x),\nonumber\\
\end{eqnarray*}
\begin{eqnarray}
&& \bar C (x) \; \longrightarrow  \bar {\cal F} (x, \vartheta) = \bar C(x) + i\,\vartheta\, \bar B_2 (x),\nonumber\\
&& {\cal B} (x) \;\longrightarrow \tilde {\cal B} (x, \vartheta) = {\cal B} (x) + \vartheta\, \bar f_1 (x),\nonumber\\
&& \bar {\cal B} (x) \; \longrightarrow {\tilde {\bar {\cal B}}} (x, \vartheta) = \bar {\cal B} (x) + \vartheta\, \bar f_2 (x),\nonumber\\
&& \rho (x) \;\; \longrightarrow {\tilde \rho} (x, \vartheta) = \rho (x) + \vartheta\, \bar f_3 (x),
\end{eqnarray}
where the coefficients of $\vartheta$  (i.e. $\bar {\cal R}_\mu, \, \bar {\cal S}_\mu, \, \bar B_1, \, \bar B_2, \, \bar f_1,\, \bar f_2,\; \bar f_3$) 
are nothing but the secondary fields which we have to  determine  by exploiting  the  ideas of ACSA to BRST formalism.
The fermionic nature of  $\vartheta$  verifies  that the secondary fields $(\bar B_1, \, \bar B_2)$ 
are bosonic in nature and $(\bar {\cal R}_\mu, \, \bar {\cal S}_\mu, \, \bar f_1,\, \bar f_2,\; \bar f_3)$ are fermionic in nature.

For the derivation of the above secondary fields, 
we use the very important  and interesting 
anti-BRST invariant quantities which are the combinations of the basic and auxiliary fields 
of the Lagrangians densities, namely:
\begin{eqnarray}
&& s_{ab} \bar {\cal B} = 0, \quad s_{ab} ({\cal D}_\mu \bar C) = 0, \quad s_{ab} (\bar C\times \bar C) = 0, 
\quad s_{ab} ({\cal A}_\mu\cdot \partial_\mu \bar {\cal B} 
 + i\, {\cal D} _\mu \bar C\cdot {\partial}^\mu C) = 0, \nonumber\\
&& s_{ab} ({\cal B}\times \bar C) = 0, \quad s_{ab} (\rho \times \bar C) = 0, \quad s_{ab} (\phi_\mu \times \bar C) = 0,\quad
s_{ab} (\bar {\cal B}\times C) = 0.
\end{eqnarray}
According to the basic principle of the ACSA to BRST formalism, above set of BRST invariant restrictions 
must be independent of the coordinate $\vartheta$ when these invariant quantities are generalized 
onto the (3, 1)-dimensional chiral super sub-manifold  as:
\begin{eqnarray}
&&{\tilde {\bar {\cal B}} }(x, \vartheta) = \bar {\cal B} (x),  \quad 
\partial_\mu \bar {\cal F} (x, \vartheta) -  g\,  B_\mu (x, \vartheta) \times \bar {\cal F} (x, \vartheta)  = \partial_\mu \bar C - g\,{\cal A}_\mu (x) \times \bar C(x), \nonumber\\
&&{\tilde {\bar {\cal F}}} (x, \vartheta) \times {{\bar {\cal F}}} (x, \vartheta) = \bar C(x)\times \bar C(x),
 \quad B_\mu (x, \vartheta)\cdot
\partial_\mu \bar {\cal B}(x, \vartheta) + i\, [\partial^\mu \bar {\cal F}(x, \vartheta)\nonumber\\
&&  - g\, B^\mu (x, \vartheta)\times \bar {\cal F}(x, \vartheta)]\cdot \partial _\mu  {\cal F} (x, \vartheta) = {\cal A}_\mu (x)\cdot \partial_\mu {\cal B} (x) + i\, \partial_\mu [\partial^\mu \bar C(x) - \, g\, ({\cal A}^\mu (x)\nonumber\\
&&\times \bar C(x))]\cdot C (x), \quad {\tilde{{\cal B}}} (x, \vartheta) \times \bar {\cal F} (x, \vartheta) = {{\cal B}} (x) \times \bar C(x), \quad 
\tilde \rho (x, \vartheta)\times  \bar {\cal F} (x, \vartheta) = \rho (x) \times \bar C(x),\nonumber\\
 && {\tilde {\bar {\cal B}}} (x, \vartheta)\times  {\cal F} (x, \vartheta) = \bar {\cal B} \times C(x), \quad 
\tilde \Phi_\mu (x, \vartheta)\times  {\cal F} (x, \vartheta) = \phi_\mu (x)\times C(x). 
\end{eqnarray}
Using the above generalized quantities, as in the similar manner to anti-chiral,  we find out the values of chiral secondary fields
of the expansions of the chiral superfields (37) as:
\begin{eqnarray}
&& \bar {\cal R}_\mu = {\cal D_\mu} \bar C, \qquad  \bar B_1 =\bar {\cal B},\qquad \bar B_2 =   -\, \frac {i\,g}{2} \,(\bar C\times \bar C),
\qquad \bar f_2 = 0\nonumber\\
&& \bar f_1 = -\, g\, ({\cal B} \times \bar C),\qquad \bar {\cal S}_\mu  = -\,g\, (\phi_\mu \times \bar C), \qquad \bar f_3 
= -\,g\, (\rho \times \bar C). 
\end{eqnarray}
We have the following super expansions of the chiral superfield after the substitutions of the above 
secondary fields  into Eq. (37); 
\begin{eqnarray}
&& {\cal A}_\mu (x) \longrightarrow \; B_\mu ^{(ab)} (x, \vartheta)  
= {\cal A}_\mu (x) + \vartheta\, [{\cal D_\mu} \bar C (x)] = 
{\cal A}_\mu (x) + \vartheta\, [s_{ab} {\cal A}_\mu (x)],\nonumber\\
&& \phi_\mu (x) \; \longrightarrow \; \tilde\Phi_\mu ^{(ab)} (x, \vartheta) 
= \phi_\mu (x) + \vartheta\, [-\,g\, (\phi_\mu (x) \times \bar C (x))]
= \phi_\mu (x) + \vartheta\, [s_{ab}\phi_\mu (x)],\nonumber\\
&& C (x) \;\; \longrightarrow  \; {\cal F} ^{(ab)} (x, \vartheta) 
= C(x) + \vartheta\, [i\, \bar{\cal B} (x)]
= C(x) + \vartheta\, [s_{ab} C(x)],  \nonumber\\
&& \bar C (x) \;\;\longrightarrow \;  \bar {\cal F} ^{(ab)} (x, \vartheta) 
= \bar C(x) + \vartheta\, [\frac {g}{2} \,(\bar C(x)\times \bar C(x))]
= \bar C(x) + \vartheta\, [s_{ab} \bar C (x)],\nonumber\\
&& {\cal B} (x) \;\;\,\longrightarrow \; \tilde {\cal B} ^{(ab)} (x, \vartheta) 
= {\cal B} (x) + \vartheta\, [-\, g\, ({\cal B}(x) \times \bar C (x))]
= {\cal B} (x) + \vartheta\, [s_{ab} {\cal B} (x)],\nonumber\\
&& \bar {\cal B} (x) \;\;\, \longrightarrow \; {\tilde {\bar {\cal B}}} ^{(ab)} (x, \vartheta) 
= \bar {\cal B} (x) + \vartheta\, [0]
= \bar {\cal B} (x) + \vartheta\, [s_{ab} \bar {\cal B} (x)],\nonumber\\
&& \rho (x) \;\;\; \longrightarrow \; {\tilde \rho} ^{(ab)} (x, \vartheta) 
= \rho (x) + \vartheta\, [-\,g\, (\rho (x) \times \bar C (x))]
= \rho (x) + \vartheta\, [s_{ab} \rho (x)],
\end{eqnarray}
where the superscript $(ab)$ on the superfields denotes the chiral superfields that have been obtained after the 
application of the anti-BRST invariant restrictions (38). Here, the coefficients of the $\vartheta$ are nothing but the anti-BRST symmetry 
transformations for various fields of Lagrangian densities which have been listed in the Eqs. (5) and (6). 
Thus, we have derived all the anti-BRST symmetry transformations for coupled 
Lagrangian densities and shown the sanctity of anti-BRST symmetry transformations  within the ambit of ACSA.

We end this subsection with the remarks that we have derived the complete set of 
(anti-) BRST symmetry transformations for Lagrangians densities listed in Eqs. (5) and (6).  
We have also established that anti-BRST symmetry ($s_{ab}$) is connected with the translational generator 
$(\partial_{\vartheta})$ with the following relationship and mapping:  
\begin{eqnarray}
\partial_{\vartheta} \, \Omega^{(ab)} (x, \vartheta) = s_{ab} \, \omega (x), \qquad s_{ab} \longleftrightarrow \partial_{\vartheta}.
\end{eqnarray}
It is clear from the above relationship  that the anti-BRST symmetry transformation
$(s_{ab})$ for the ordinary generic field $\omega (x)$ gives the translation 
of the generalized generic chiral superfield $[\Omega^{(ab)} (x, \vartheta)]$ 
onto (3, 1)-dimensional super sub-manifold  along the $\vartheta$-direction.

\section{Nilpotency and Absolute Anticommutativity  Properties of  Conserved (Anti-)BRST Charges: ACSA}

\noindent
In this section, we discuss about the off-shell nilpotency and absolute anticommutativity
properties of the (anti-)BRST conserved charges within the framework of ACSA to BRST formalism.
For the proof of the nilpotency properties of the conserved (anti-)BRST charges [${\cal Q}_{(a)b}$], we express the charges 
in terms of the anti-chiral and chiral superfields as: 
\begin{eqnarray}
{\cal Q}_{b} &=& \frac {\partial}{\partial\bar\vartheta}\int\; d^{2}x\;\Big[{\cal B}^{(b)} (x, \bar\vartheta)\cdot B_0^{(b)}(x,\bar\vartheta) + 
 i\;{\dot {\bar {\cal F}}}^{(b)}(x,\bar\vartheta)\;\cdot {\cal F}^{(b)}(x,\bar\vartheta)  \Big]\nonumber\\
&\equiv & \int\; d\bar\vartheta\;\int\; d^{2}x\;\Big[{\cal B}^{(b)} (x, \bar\vartheta)\cdot B_0^{(b)}(x,\bar\vartheta) + 
 i\;\dot {\bar F}^{(b)}(x,\bar\vartheta)\;\cdot {\cal F}^{(b)}(x,\bar\vartheta)\Big],\nonumber\\
{\cal Q}_{ab} &=& \frac {\partial}{\partial\vartheta}\int\; d^{2}x\;\Big[\bar {\cal B}^{(b)}  (x, \vartheta)\cdot B_0^{(ab)}(x, \vartheta) 
-  i\;\bar {\cal F}^{(ab)}(x,\vartheta)\;\cdot\dot {\cal F}^{(ab)}(x,\vartheta)  \Big]\nonumber\\
&\equiv & \int\; d\vartheta\;\int\; d^{2}x\;\Big[ \bar {\cal B}^{(b)}  (x, \vartheta)\cdot B_0^{(ab)}(x, \vartheta) 
-  i\;\bar {\cal F}^{(ab)}(x,\vartheta)\;\cdot\dot {\cal F}^{(ab)}(x,\vartheta)\Big].
\end{eqnarray}
It is now crystal clear that we have the following interesting relationships:
\begin{eqnarray}
&&\partial_{\bar\vartheta}\; {\cal Q}_b = 0 \;\,~\Longleftrightarrow \, \partial_{\bar\vartheta}^2 = 0,
~~\;\, s_b\;{\cal Q}_b \;\;= -\,i\,{\{{\cal Q}_b, {\cal Q}_b}\} = 0\;\;\;\, \Longleftrightarrow \, s_b^2 = 0,\nonumber\\
&&\partial_\vartheta \;{\cal Q}_{ab} = 0\; \, \Longleftrightarrow \, \partial_\vartheta^2 = 0,
~~\, s_{ab}\;{\cal Q}_{ab} = -\,i\,{\{{\cal Q}_{ab}, {\cal Q}_{ab}}\} = 0\, \,\Longleftrightarrow\, s_{ab}^2 = 0.
\end{eqnarray}
In other words, it is clear that the nilpotency of the conserved (anti-)BRST charges 
(i.e. ${\cal Q}_{b}^2 = {\cal Q}_{ab}^2 = 0)$ is deeply connected with the nilpotency 
($\partial_{\bar\vartheta}^2 = \partial_{\vartheta}^2 = 0$) of the translational generators $(\partial_{\bar\vartheta}, \partial_{\vartheta})$, respectively, 
along Grassmannian directions and the nilpotency ($s_{b}^2 = s_{ab}^2 =  0$) of the (anti-)BRST symmetry transformations $(s_{(a)b})$, too.

We now focus on the  proof of  absolute anticommutativity property of the nilpotent conserved (anti-)BRST charges within the realm  of 
ACSA to BRST formalism where the CF-condition play major role. The conserved (anti-)BRST charges can be expressed in terms of the (anti-)chiral 
superfields of the (3, 1)-dimensional super sub-manifold as:   
\begin{eqnarray}
 {\cal Q}_b     & = & \frac {\partial}{\partial\vartheta}\;\Big [ \int\; d^{2}x\;\Big\{i\; {\cal F}^{(ab)}(x, \vartheta)\cdot \dot {\cal F}^{(ab)}
(x,\vartheta)\nonumber\\
 & - &  \frac {1}{2} \;{\cal F}^{(ab)}(x, \vartheta)\cdot \Big[{B}^{(ab)}_0(x, \vartheta)\times {\cal F}^{(ab)}(x, \vartheta)\Big]\Big\}\Big]\nonumber\\
  & \equiv  &   \int\; d\,\vartheta\;\int\; d^{2}x\;\Big [ \Big\{i\; {\cal F}^{(ab)}(x, \vartheta)\cdot \dot {\cal F}^{(ab)}(x, \vartheta)\nonumber\\
 & - &  \frac {1}{2} \;{\cal F}^{(ab)}(x, \vartheta)\cdot \Big[B^{(ab)}_0(x, \vartheta)\times {\cal F}^{(ab)}(x, \vartheta)\Big]\Big\}\Big], \nonumber\\
{\cal Q}_{ab}   & = & \frac {\partial}{\partial\bar\vartheta}\;\Big [ \int\; d^{2}x\;\Big\{-i\;\;\bar  {\cal F}^{(b)}(x, \bar\vartheta)\cdot 
{\dot {{\bar {\cal F}}}}^{(b)} (x, \bar\vartheta)\nonumber\\
 & + &  \frac {1}{2}\;\bar {\cal F}^{(b)}(x,\bar\vartheta)\cdot (B^{(b)}_0(x,\bar\vartheta)\times {\bar {\cal F}}^{(b)}(x,\bar\vartheta))\Big\}\Big]\nonumber\\
 & \equiv  &   \int\; d\,\bar\vartheta\; \int\; d^{2}x\;\Big [-i\;\;\bar  {\cal F}^{(b)}(x,\bar\vartheta)\cdot {\dot {\bar {\cal F}}}^{(b)}(x,\bar\vartheta)\nonumber\\
 & + &  \frac {1}{2}\;\bar {\cal F}^{(b)}(x,\bar\vartheta)\cdot (B^{(b)}_0(x,\bar\vartheta)\times {\bar {\cal F}}^{(b)}(x,\bar\vartheta))\Big].
\end{eqnarray}
From the above expressions, it is clear that:
\begin{eqnarray}
&&\partial_\vartheta {\cal Q}_b  = 0\quad\;\Longleftrightarrow\quad \partial_\vartheta^2 = 0\quad\Longleftrightarrow \quad s_{ab} {\cal Q}_b = - i\;{\{{\cal Q}_b, {\cal Q}_{ab}}\} = 0, \nonumber\\
&&\partial_{\bar\vartheta} {\cal Q}_{ab} = 0\quad \Longleftrightarrow\quad\partial_{\bar\vartheta}^2 = 0\quad \Longleftrightarrow \quad 
s_b  {\cal Q}_{ab} = - \;i\;{\{{\cal Q}_{ab}, {\cal Q}_b}\} = 0.
\end{eqnarray}
Thus, we note that the absolute anticommutativity of the BRST charge $({\cal Q}_b)$ {\it with} the anti-BRST charge $({\cal Q}_{ab})$ 
is deeply connected with the nilpotency $(\partial_\vartheta^2 = 0)$ of the  translational generator $(\partial_\vartheta)$ along
the $\vartheta$-direction. Similarly, absolute anticommutativity of the anti-BRST charge with BRST charge is deeply connected with the 
nilpotency of the  $(\partial_{\bar\vartheta} ^2 = 0)$  of the translational generator  $(\partial_{\bar \vartheta})$ along
the $\bar \vartheta$-direction of the (3, 1)-dimensional  super sub-manifold of the general (3, 2)-dimensional
supermanifold.

\section{Invariances of Lagrangian Densities: ACSA}

In this section, we capture the (anti-)BRST invariances of the Lagrangian densities
${\cal L}_{\cal B}$ and ${\cal L}_{\bar {\cal B}}$ [Eqs. (5) and (6)] within the  realm of ACSA to BRST 
formalism. Towards this aim in mind, foremost, we generalize the ordinary Lagrangian densities  ${\cal L}_{\cal B}$ and ${\cal L}_{\bar {\cal B}}$
to (anti-)chiral super Lagrangian densities  $\tilde {\cal L}^{(ac)}_{\cal B}$  and $\tilde {\cal L}^{(c)}_{\bar {\cal B}}$ 
onto the (3, 1)-dimensional super sub-manifold  as:
\begin{eqnarray*}
{\cal L}_{\cal B}  &\longrightarrow & \tilde {\cal L}^{(ac)}_{\cal B}  =  -\,\frac {1}{4}\, \tilde {\cal F}^{\mu\nu( ac)}
(x, \bar\vartheta)\cdot\,\tilde {\cal F}_{\mu\nu}^{( ac)}(x, \bar\vartheta)\nonumber\\
& - & \frac {1}{4}\; \big[\tilde {\cal {G}}^{\mu\nu (ac)} (x, \bar\vartheta) 
+ g\; \tilde {\cal F}^{\mu\nu (ac)} (x, \bar\vartheta)  \times \tilde \rho ^{(b)} 
(x, \bar\vartheta) \big] \cdot \big [\tilde  {\cal {G}}_{\mu\nu}^{(ac)} (x, \bar\vartheta) \nonumber\\
& + &  g \;\tilde {\cal F}_{\mu\nu}^{(ac)} (x, \bar\vartheta)  \times \tilde\rho^{(b)} (x, \bar\vartheta) \big] 
+ \frac {m}{2}\;\varepsilon^{\mu\nu\eta} \; {\cal F}_{\mu\nu}^{(ac)} (x, \bar\vartheta)  \cdot \tilde \Phi_\eta ^{(b)}  (x, \bar\vartheta)\nonumber\\
\end{eqnarray*}
\begin{eqnarray}
& + & \tilde {\cal B}^{(b)} (x, \bar\vartheta) \cdot \big [\partial^\mu B_\mu ^{(b)} (x, \bar\vartheta)  \big] +  \frac {1}{2}\;\Big[\tilde {\cal B} ^{(b)}(x, \bar\theta)\cdot \tilde {\cal B} ^{(b)} (x, \bar\theta) + {\tilde {\bar  {\cal B}}}^{(b)} (x, \bar\vartheta) \cdot {\tilde {\bar  {\cal B}}}^{(b)}(x, \bar\vartheta)\Big]\nonumber\\
 & - &  i \;\partial_{\mu}{\bar {\cal F}}^{(b)}(x, \bar\vartheta)\cdot\partial^{\mu} {\cal F}^{(b)}(x, \bar\vartheta)+ \partial_\mu{\bar {\cal F}}^{(b)}(x, \bar\vartheta)\cdot
\Big[B^{\mu(b)}(x,\bar\vartheta)\times {\cal F}^{(b)}(x,\bar\vartheta)\Big],
\nonumber\\
{\cal L}_{\cal {\bar B}}  &\longrightarrow & \tilde {\cal L}^{(c)}_{\cal {\bar B}}  =  -\,\frac {1}{4}\, \tilde {\cal F}^{\mu\nu(c)}
(x, \vartheta)\cdot\,\tilde {\cal F}_{\mu\nu}^{(c)}(x, \vartheta)\nonumber\\
& - & \frac {1}{4}\; \big[\tilde {{\cal {G}}}^{\mu\nu (c)} (x, \vartheta) 
+ g\; \tilde {\cal F}^{\mu\nu (c)} (x, \vartheta)  \times \tilde \rho ^{(ab)} 
(x, \vartheta) \big] \cdot \big [\tilde  {\cal {G}}_{\mu\nu}^{(c)} (x, \vartheta) \nonumber\\
& + &  g \;\tilde {\cal F}_{\mu\nu}^{(c)} (x, \vartheta)  \times \tilde\rho^{(ab)} (x, \vartheta) \big] 
 +  \frac {m}{2}\;\varepsilon^{\mu\nu\eta} \; {\cal F}_{\mu\nu}^{(ac)} (x, \vartheta)  \cdot \tilde \Phi^{(ab)}_\eta (x, \vartheta)\nonumber\\ 
 & - &  {\tilde {\bar {\cal B}}}^{(ab)} (x, \vartheta) \cdot \big [\partial^\mu B_\mu ^{(ab)} (x, \bar\vartheta)  \big] 
 +   \frac {1}{2}\;\Big[\tilde {\cal B}^{(ab)} (x, \vartheta)\cdot \tilde {\cal B}^{(ab)}  (x, \vartheta) 
+ {\tilde {\bar  {\cal B}}}^{(ab)} (x, \vartheta) \cdot {\tilde {\bar  {\cal B}}}^{(ab)}(x, \vartheta)\Big] \nonumber\\
 & - &  i \;\partial_{\mu}{\bar {\cal F}}^{(ab)}(x, \vartheta)\cdot\partial^{\mu} {\cal F}^{(ab)}(x, \vartheta) 
+  \Big[B^{\mu(ab)}(x, \vartheta)\times \bar {\cal F}^{(ab)}(x, \vartheta)\Big]\cdot\partial_\mu{{\cal F}}^{(ab)}(x, \vartheta),
\end{eqnarray}
where the superscripts $(ac)$,  $(c)$ on the super Lagrangian densities and field  strength tensors
denote the anti-chiral and   chiral super Lagrangian densities and super field strength tensors,
respectively,  which are generalized on the (3, 1)-dimensional super submanifolds.
It is straightforward to check that after the application of the translational generators
 $\partial_{\bar\vartheta}$ and $\partial_{\vartheta}$ on the above super Lagrangian densities
 $\tilde {\cal L}^{(ac)}_{\cal B}$ and  $\tilde {\cal L}^{(c)}_{\bar {\cal B}}$, respectively, we have the following 
\begin{eqnarray}
&&\frac {\partial}{\partial\bar\vartheta} \;\Big [\tilde {\cal L}^{(ac)}_{\cal B}\Big] = \partial_\mu 
\;[{\cal B}\cdot {\cal D}^\mu C]\qquad\Longleftrightarrow\quad s_b \;{\cal L}_{\cal B} = \partial_\mu [{\cal B}\cdot {\cal D}^\mu C],
\nonumber\\
&&\frac {\partial}{\partial\vartheta} \;\Big [\tilde {\cal L}^{(c)}_{\bar {\cal B}}\Big] = \partial_\mu 
\;[ -\; \bar {\cal B} \cdot {\cal D}^\mu \bar C]\quad\;\,\Longleftrightarrow \quad s_{ab} \;{\cal L}_{\bar {\cal B}}= -\;\partial_\mu [\bar {\cal B}\cdot {\cal D}^\mu\bar  C].
\end{eqnarray}
The above relationships  establish that the invariances of the Lagrangian densities  within the realm of ACSA to BRST formalism 
which demonstrate the sanctity of the (anti-)BRST invariance [cf. Eq. (12)] of the Lagrangian densities. 
It should also be noted that (anti-) BRST symmetries  ($s_{(a)b}$) are connected with the translational generators 
$(\partial_{\bar\vartheta}, \,\partial_{\vartheta})$.

To capture the anti-BRST invariance the of Lagrangian density ${\cal L}_{\cal {B}}$ and BRST invariance of the 
Lagrangian density ${\cal L}_{\cal {\bar B}}$  within the framework of (anti-)chiral superfields formalism, first of all,  
we generalize ordinary Lagrangian densities ${\cal L}_{\cal {\bar B}}$ and ${\cal L}_{\cal {B}}$ onto the (3, 1)-dimensional super sub-manifold
of the most general (3, 2)-dimensional supermanifold as  
\begin{eqnarray*}
{\cal L}_{\cal B}  &\longrightarrow & \tilde {\cal L}^{(c)}_{\cal B}  =  -\,\frac {1}{4}\, \tilde {\cal F}^{\mu\nu(c)}
(x, \vartheta)\cdot\,\tilde {\cal F}_{\mu\nu}^{( c)}(x, \vartheta)\nonumber\\
& - & \frac {1}{4}\; \big[\tilde {{\cal {G}}}^{\mu\nu (c)} (x, \vartheta) 
+ g\; \tilde {\cal F}^{\mu\nu (c)} (x, \vartheta)  \times \tilde \rho ^{(ab)} 
(x, \vartheta) \big] \cdot \big [\tilde  {\cal {G}}_{\mu\nu}^{(c)} (x, \vartheta) \nonumber\\
& + &  g \;\tilde {\cal F}_{\mu\nu}^{(c)} (x, \vartheta)  \times \tilde\rho^{(ab)} (x, \vartheta) \big]
+ \frac {m}{2}\;\varepsilon^{\mu\nu\eta} \; {\cal F}_{\mu\nu}^{(c)} (x, \vartheta)  \cdot \tilde \Phi_\eta ^{(ab)}  (x, \vartheta)\nonumber\\
& + & \tilde {\cal B}^{(ab)} (x, \vartheta) \cdot \big [\partial^\mu B_\mu ^{(ab)} (x, \vartheta)  \big] +  \frac {1}{2}\;\Big[\tilde {\cal B} ^{(ab)}(x, \bar\theta)\cdot \tilde {\cal B} ^{(ab)} (x, \bar\theta) + {\tilde {\bar  {\cal B}}}^{(ab)} (x, \vartheta) \cdot {\tilde {\bar  {\cal B}}}^{(ab)}
(x, \vartheta)\Big]\nonumber\\
 & - &  i \;\partial_{\mu}{\bar {\cal F}}^{(ab)}(x, \vartheta)\cdot\partial^{\mu} {\cal F}^{(ab)}(x, \vartheta)+ \partial_\mu{\bar {\cal F}}^{(ab)}(x, \vartheta)\cdot
\Big[B^{\mu(b)}(x,\vartheta)\times {\cal F}^{(ab)}(x,\vartheta)\Big],
\nonumber\\
\end{eqnarray*}
\begin{eqnarray}
{\cal L}_{\cal {\bar B}}  &\longrightarrow & \tilde {\cal L}^{(ac)}_{\cal {\bar B}}  =  -\,\frac {1}{4}\, \tilde {\cal F}^{\mu\nu (ac)}
(x, \bar\vartheta)\cdot\,\tilde {\cal F}_{\mu\nu}^{(ac)}(x, \bar\vartheta)\nonumber\\
& - & \frac {1}{4}\; \big[\tilde {{\cal {G}}}^{\mu\nu (ac)} (x, \bar\vartheta) 
+ g\; \tilde {\cal F}^{\mu\nu (ac)} (x, \bar\vartheta)  \times \tilde \rho ^{(b)} 
(x, \bar\vartheta) \big] \cdot \big [\tilde  {\cal {G}}_{\mu\nu}^{(ac)} (x, \bar\vartheta) \nonumber\\
& + &  g \;\tilde {\cal F}_{\mu\nu}^{(ac)} (x, \bar\vartheta)  \times \tilde\rho^{(b)} (x, \bar\vartheta) \big] 
 +  \frac {m}{2}\;\varepsilon^{\mu\nu\eta} \; {\cal F}_{\mu\nu}^{(ac)} (x, \bar\vartheta)  \cdot \tilde \Phi^{(b)}_\eta (x, \bar\vartheta)\nonumber\\ 
 & - &  {\tilde {\bar {\cal B}}}^{(b)} (x, \bar\vartheta) \cdot \big [\partial^\mu B_\mu ^{(b)} (x, \bar\vartheta)  \big] 
 +   \frac {1}{2}\;\Big[\tilde {\cal B}^{(b)} (x, \bar\vartheta)\cdot \tilde {\cal B}^{(b)}  (x, \bar\vartheta) 
+ {\tilde {\bar  {\cal B}}}^{(b)} (x, \bar\vartheta) \cdot {\tilde {\bar  {\cal B}}}^{(b)}(x, \bar\vartheta)\Big] \nonumber\\
 & - &  i \;\partial_{\mu}{\bar {\cal F}}^{(b)}(x, \bar\vartheta)\cdot\partial^{\mu} {\cal F}^{(b)}(x, \bar\vartheta) 
+ \Big[B^{\mu(b)}(x, \bar\vartheta)\times \bar {\cal F}^{(b)}(x, \bar\vartheta)\Big]\cdot \partial_\mu{{\cal F}}^{(b)}(x, \bar\vartheta),
\end{eqnarray}
where superscripts  $(c)$ and $(ac)$ have been explained  earlier in Eq. (47). 
We apply the translational generators $({\partial}_{\vartheta}, {\partial}_{\bar\vartheta})$ on the 
the above super Lagrangian densities $[\tilde {\cal L}^{(c)}_{\cal {B}}, \tilde{\cal L}^{(ac)}_{\cal {\bar B}}]$, respectively, as follows: 
\begin{eqnarray}
\frac {\partial}{\partial\vartheta} \, \Big[\tilde {\cal L}^{(c)}_{\cal B}\Big] & = & 
-\;\partial_\mu \big[\bar {\cal B}+ (\bar C\times C)\big]\cdot \partial^\mu\bar C
 - {\cal D}_\mu\, [{\cal B} + \bar {\cal B} + i\,g\,(\bar C\times C)]\cdot \partial^\mu\bar C,\nonumber\\
\frac {\partial}{\partial{\bar\vartheta}}\, \Big[\tilde {\cal L}^{(ac)}_{\bar {\cal B}}\Big]  &=& 
~~ \partial_\mu \big[{\cal B} + (\bar C\times C)\big]\cdot \partial^\mu C
+ {\cal D}_\mu\, [{\cal B} + \bar {\cal B} + i\,g\,(\bar C\times C)]\cdot \partial^\mu C.
\end{eqnarray}
The above relationships lead to the invariances of Lagrangian densities if and only if  CF-condition
[i.e. ${\cal B} + \bar {\cal B} + i\,g\,(\bar C\times C) = 0]$ is satisfied. Therefore, we prove the sanctity of the
invariance of  coupled (equivalent) Lagrangian densities [cf. Eq. (12)], as in ordinary spacetime, 
within the  realm of the ACSA to BRST formalism.\\

\section{Conclusions}

In the present investigation, first of all, we have talked about the Yang-Mills (YM) and non-Yang-Mills (NYM) gauge symmetry transformations  for (2 + 1)-dimensional (3D) 
non-Abelian massive 1-form Jackiw-Pi model (parity conserving odd-dimensional theory). We have discussed the construction  of 
coupled (but equivalent) Lagrangian densities, BRST and anti-BRST symmetry transformations for various basic and auxiliary  fields present 
in the Lagrangian densities, and invariances of Lagrangian densities under (anti-)BRST symmetry transformations  in ordinary spacetime (cf. Sec. 3).
 We have also discussed the Noether conserved (anti-)BRST currents as well as charges and shown the conservation law using the Euler Lagrange equation of motion.
Along with these discussions, we have also examined  the nilpotency and absolute anticommutativity properties of the 
(anti-)BRST conserved charges in the 3D ordinary  spacetime. One of the key signatures of any non-Abelian gauge theory
is  the existence of the  Curci-Ferrari (CF)-condition which is observed, for the present model,  through the various mathematical  
techniques (i) equivalence of both the Lagrangian densities (i.e. ${\cal L}_{\cal B} \, \equiv \,  {\cal L}_{\bar {\cal B}} $)
(ii) absolute anticommutativity of (anti-)BRST symmetries for the various fields (iii) invariances of coupled Lagrangian 
densities under the off-shell  nilpotent (anti-)BRST symmetry transformations, and (iv) absolute anticommutativity property of
the conserved (anti-)BRST charges. Moreover, CF-condition has played a crucial  role in the deduction of coupled
(but equivalent) Lagrangian densities.

We have discussed, for the first time, all the above properties (of ordinary spacetime) 
 within the realm of (anti-)chiral superfield approach (ACSA) to BRST formalism  and shown the obvious 
connections of (anti-)BRST symmetry transformations with the Grassmannian derivatives (i.e. $s_b \longleftrightarrow \partial_{\bar\vartheta}, 
\; s_{ab} \longleftrightarrow \partial_\vartheta$).
One of the important and novel result of the ACSA to BRST formalism for the JP model  is that absolute anticommutativity property of the (anti-)BRST 
charges are satisfied even though  we have taken only one Grassmannian coordinate in the superfields  whereas in Bonora-Tonin (BT)
superfield formalism both the Grassmannian coordinate are taken into account for the proof of absolute anticommutativity of charges.
We have observed that there is a deep connection between the nilpotency of the translational generator $\partial_{\bar\vartheta}$ and 
anticommutativity of the BRST conserved charge ${\cal Q}_{b}$ with anti-BRST charge  ${\cal Q}_{ab}$.
Whereas there is deep connection between the nilpotency of translational generator $\partial_\vartheta$
and anticommutativity of the anti-BRST charge ${\cal Q}_{ab}$ with BRST charge ${\cal Q}_b$. 
We have also demonstrated the nilpotency properties of (anti-)BRST conserved charges within the real of ACSA to BRST formalism where 
nilpotency of the translational generators ($\partial_{\vartheta}, \partial_{\bar\vartheta}$) 
is deeply connected with the nilpotency of the conserved charges (${\cal Q}_{ab}, {\cal Q}_{b}$) along the
$(\vartheta, \bar\vartheta)$-directions of the (3, 1)-dimensional (anti-)chiral super sub-manifold of the general (3, 2)-dimensional 
supermanifold,  respectively [cf. Eq. (44)]. We have also shown the invariance of the  coupled (but equivalent) Lagrangian  densities 
where the emergence of the CF-condition is verified within the ambit of the ACSA to BRST formalism.

We would extend our standard techniques of (anti-)chiral superfield approach to BRST formalism 
for the various BRST-quantized models and theories. The techniques of the ACSA with modified 
Bonora-Tonin superfield approach (MBTSA) would be very interesting to discuss the various  
reparameterization invariant  models. We also plan to discuss the  massive
Abelian 3-form gauge theory in 6D within the realm of ACSA where the possible candidate
of dark matter and dark energy would be discussed within the realm of BRST formalism.\\

\noindent
{\bf\large Data Availability}\vskip 0.2cm

\noindent
No data were used to support this study.\\

\noindent
{\bf\large Conflicts of Interest}\vskip 0.2cm

\noindent
The author declares that there is no conflicts of interest.\\ \vskip 0.2cm

\noindent
{\large\bf Acknowledgments}\\

\noindent
The present investigation has been carried out under the
DST-INSPIRE fellowship (Govt. of India) awarded to the author. The author  of this paper  expresses their gratefulness to
the above national funding agency  as well as  A. K. Rao for a careful reading of the manuscript.
Fruitful and enlightening comments by our esteemed
Reviewer is thankfully acknowledged.

\end{document}